# Joule heating effect on Peristaltic transport of water-based copper oxide over an asymmetric channel in presence of slip effects

Asha. S. K[1] and Sunitha G[2]

[1,2] Department of Mathematics, Karnatak University, Dharwad-580003, India
ashask@kud.ac.in [1] and sunithag643@gmail.com[2]

**Abstract:** This research paper is intended to study the nanoparticles shapes on peristalsis of Casson fluid in a channel flow by considering Joule heating effects. The study of such flows has various relevance applications in biomedical engineering and industries. Slip conditions is maintained at the velocity, temperature and nanoparticle concentration. The problem is modeled in terms of partial differential equations with suitable slip boundary conditions and then computed by using semi analytical technique known as Homotopy Analysis Method with Mathematica software. The influences of nanoparticles shape on velocity, pressure, temperature and nanoparticle concentration distributions are discussed with the help of graphs. Here, increase in Joule heating parameter enhance the fluid temperature and nanoparticle concentration decreases with an increasing Joule heating and also we observed that the different shapes has different thermal conductivity, but the lamina shaped nanoparticles have high thermal conductivity as compared to needle shaped nanoparticles.

**Nomenclature**

| | |
|---|---|
| $\hat{X}, \hat{Y}$ | Cartesian coordinates |
| $\hat{t}$ | Time |
| $\hat{h}_1$ | Right wall |
| $\beta_2$ | Thermal slip parameter |
| $\hat{u}, \hat{v}$ | Velocity components |
| Re | Reynolds number |
| $\beta$ | Casson fluid parameter |
| $\beta_1$ | Velocity slip parameter |
| $Sr$ | Soret number |
| $\hat{p}$ | Pressure |
| $\hat{h}_2$ | Left wall |

| Symbol | Description |
|---|---|
| $\hat{T}$ | Temperature |
| $Br$ | Nanoparticle Grashof number |
| $\hat{\varphi}$ | Nanoparticle concentration |
| $\beta_3$ | Concentration slip parameter |
| $Nt$ | thermophoresis diffusion parameter |
| $T_m$ | Fluid mean temperature |
| $K_{nf}$ | Effective thermal conductivity of the nanofluid |
| $\rho_f$ | Base fluid density |
| $Nb$ | Brownian motion parameter |
| $Gr$ | Thermal Grashof number |
| $\rho_s$ | Copper (II) oxide density |
| $(\rho C_p)_{nf}$ | Heat capacitance |
| $\rho_{nf}$ | Effective density |
| $\phi$ | Solid nanoparticle volume fraction |
| $\mu_{nf}$ | Effective dynamic viscosity |
| $g$ | Acceleration due to gravity |
| $\alpha_{nf}$ | Effective thermal diffusivity |
| $Q_0$ | Joule heating effect |
| $(\rho\beta)_{nf}$ | the coefficient of thermal expansion |
| $\psi$ | Stream function. |

# 1. Introduction

The Analysis of peristaltic transport has great importance in biology and biomedicine. The biological applications are the vasomotion of small blood vessels such as venules, arterioles and capillaries, the process of swallowing of food through the esophagus and urine transport through the urethra. Latham [5] was first to investigate the peristaltic transport. Newtonian fluids in the wave frame and laboratory frame was studied by Shapiro et al [6] and Fung and Yih [7]. Some researchers [8-11] have investigated the peristalsis through different channel flows by considering Newtonian and non- Newtonian fluids.

The joule heating is referred to as electrically resistance heating or ohmic heating because of its relationship with ohm's law. Electric stoves, cartridge heaters, soldering irons, electric fuses and incandescent light (glows when the filament is heated by joule heating) are the important practical applications of Joule heating. Recent attempts on peristaltic transport with joule heating effect can be seen in references [12-15].

The nanofluids are a new type of fluids carried a very tiny number of nanoparticles are fixed effect in a carrier liquid. The thermal conductivity of the fluids is very low compared to solids. The term "nanofluid" was first introduced by Choi [16]. It indicates the engineered colloids composing the nanoparticles dispersed in a base fluid. The practical investigation has shown that the influence of nanoparticles has increasing the thermal conductivity depends on size as well as shape of the nanoparticles. The different shapes of the nanoparticles are Needle, Rods, Platelet, Spheres, Lamina and Bricks. Further analysis could be seen through references [17-20].

Motivated from the above discussion, the aim of the current research work is to analyze the influences of thermal radiation, magnetic field and porous medium on peristalsis of Casson fluid through channel flow under slip conditions. To the best of our knowledge, the discussion of the different nanoparticles shape effects of peristaltic flow with the Casson fluid, this has never been no attempted before. Keeping pure water as the base fluid and Copper oxide as a selected nanoparticle in the peristaltic transportation has biological and industrial appliances. The basic governing equations have been simplified with infinite wavelength and low Reynolds number assumptions. The differential equations are simplified analytically by using Homotopy Analysis Method (HAM) [21-22]. The expressions for velocity, thermal energy and nanoparticle concentration have been computed. Analytical solutions that show the different physical parameters are shown in the graphical form.

## 2. Formulation of the problem

The geometry of model has considered the peristaltic transport of uniform thickness $d_1 + d_2$ on asymmetric channel. The sinusoidal travelling of the small amplitude waves $a_1$ and $b_1$ propagates the speed of the channel walls and s is the constant speed of the channel flow. The geometry of fluid model is defined as

$$\hat{Y} = \hat{h}_1 = a_1 \cos\left[(\hat{X} - s\hat{t})\frac{2\pi}{\lambda}\right] + d_1, \tag{1}$$

$$\hat{Y} = \hat{h}_2 = -b_1 \cos\left[(\hat{X} - s\hat{t})\frac{2\pi}{\lambda} + \omega\right] - d_2, \tag{2}$$

here $a_1$ and $b_1$ are the amplitudes of the channel flow, $\lambda$ is wavelength of channel walls, $\hat{t}$ is time, $\omega$ is the phase difference, we considering $(\hat{X}, \hat{Y})$ are the coordinate system, where $\hat{X}$ and $\hat{Y}$ are perpendicular to each other. Here $d_2$ and $d_1$ are satisfies the below condition.

$$a_1^2 + 2a_1 b_1 \cos\omega + b_1^2 \leq (d_2^2 + d_1^2). \tag{3}$$

The Rheology of equations for an isotropic and incompressible flow of a Casson fluid can be expressed as follows

$$\tau_{ij} = \left(\mu_\beta + \frac{\rho_y}{\sqrt{2\Pi}}\right) 2e_{ij} \quad \text{when} \quad \Pi \succ \Pi_C, \tag{4}$$

$$\tau_{ij} = \left(\mu_\beta + \frac{\rho_y}{\sqrt{2\Pi_C}}\right) 2e_{ij} \quad \text{when} \quad \Pi \prec \Pi_C, \tag{5}$$

where $\tau_{ij}$ is the stress tensor of the fluid with (i, j)$^{th}$ components, $\Pi = e_{ij} e_{ij}$, $e_{ij}$ is the deformation rate of (i, j)$^{th}$ component, $\rho_y$ is the yield stress of the fluid, $\mu_\beta$ is dynamic plastic viscosity of the viscous fluid, $\Pi_C$ is the critical value of the product based on viscous fluid.

Vector form of the velocity field $v$ is given as,

$$V = \left(\hat{U}(\hat{x}, \hat{y}, \hat{t}), \hat{V}(\hat{x}, \hat{y}, \hat{t})\right). \tag{6}$$

where $\hat{U}(\hat{x}, \hat{y}, \hat{t})$ and $\hat{V}(\hat{x}, \hat{y}, \hat{t})$ are the velocity components.

The dimensional governing equations for two-dimensional cartesian coordinate system can be expressed as

$$\frac{\partial \hat{U}}{\partial \hat{X}}+\frac{\partial \hat{V}}{\partial \hat{Y}}=0, \tag{7}$$

$$\rho_{nf}\left(\frac{\partial \hat{U}}{\partial \hat{t}}+\hat{U}\frac{\partial \hat{U}}{\partial \hat{X}}+\hat{V}\frac{\partial \hat{U}}{\partial \hat{Y}}\right)=-\frac{\partial \hat{p}}{\partial \hat{X}}+\mu_{nf}\left(1+\frac{1}{\beta}\right)\left(\frac{\partial^2 \hat{U}}{\partial \hat{X}^2}+\frac{\partial^2 \hat{U}}{\partial \hat{Y}^2}\right)+g(\rho\beta_T)_{nf}(\hat{T}-\hat{T}_0) \\ +g(\rho\beta_C)_{nf}(\hat{C}-\hat{C}_0), \tag{8}$$

$$\rho_{nf}\left(\frac{\partial \hat{V}}{\partial \hat{t}}+\hat{U}\frac{\partial \hat{V}}{\partial \hat{X}}+\hat{V}\frac{\partial \hat{V}}{\partial \hat{Y}}\right)=-\frac{\partial \hat{p}}{\partial \hat{Y}}+\mu_{nf}\left(1+\frac{1}{\beta}\right)\left(\frac{\partial^2 \hat{V}}{\partial \hat{X}^2}+\frac{\partial^2 \hat{V}}{\partial \hat{Y}^2}\right), \tag{9}$$

$$(\rho c_P)_{nf}\left(\frac{\partial \hat{T}}{\partial \hat{t}}+\hat{U}\frac{\partial \hat{T}}{\partial \hat{X}}+\hat{V}\frac{\partial \hat{T}}{\partial \hat{Y}}\right)=k_{nf}\left(\frac{\partial^2 \hat{T}}{\partial \hat{X}^2}+\frac{\partial^2 \hat{T}}{\partial \hat{Y}^2}\right)+D_{nf}k_T\left(\frac{\partial^2 \hat{C}}{\partial \hat{X}^2}+\frac{\partial^2 \hat{C}}{\partial \hat{Y}^2}\right)+Q_0, \tag{10}$$

$$\frac{\partial \hat{C}}{\partial \hat{t}}+\hat{U}\frac{\partial \hat{C}}{\partial \hat{X}}+\hat{V}\frac{\partial \hat{C}}{\partial \hat{Y}}=D_{nf}\left(\frac{\partial^2 \hat{C}}{\partial \hat{X}^2}+\frac{\partial^2 \hat{C}}{\partial \hat{Y}^2}\right)+\frac{D_{nf}k_T}{T_m}\left(\frac{\partial^2 \hat{T}}{\partial \hat{X}^2}+\frac{\partial^2 \hat{T}}{\partial \hat{Y}^2}\right), \tag{11}$$

where $\hat{U}$ is the velocity component in $\hat{X}$ coordinate and $\hat{V}$ is the velocity component in $\hat{Y}$ coordinate, $\sigma_{nf}$ signifies the electrical conductivity, $\beta_T$ and $\beta_C$ are the coefficients of thermal expansion and solutal expansion, $\hat{C}$ is the concentration of the nanoparticle, $D_{nf}$ is the thermal diffusivity and $T_m$ fluid mean temperature.

In equations (8)-(11), the heat capacitance $(\rho C_P)_{nf}$, effective dynamic viscosity $\mu_{nf}$, effective density $\rho_{nf}$, effective thermal conductivity of the nanofluid is $k_{nf}$ and the effective of thermal diffusivity $\alpha_{nf}$ are defined as

$$(\rho C_P)_{nf}=(1-\phi)(\rho C_P)_f+\phi(\rho C_P)_S, \tag{12}$$

$$\mu_{nf}=\frac{\mu_f}{(1-\phi)^{2.5}}, \tag{13}$$

$$\rho_{nf}=(1-\phi)\rho_f+\phi\rho_S, \tag{14}$$

$$\alpha_{nf}=\frac{k_{nf}}{(\rho C_P)_{nf}}, \tag{15}$$

here $\beta_f$ and $\beta_S$ are the coefficients of thermal expansions of base fluid and nanoparticle, $(C_P)_S$ and $(C_P)_f$ are the heat capacities at a certain pressure and the Maxwell-Garnett [23] and further established Hamilton and Crosser model [24] to taken into account of random motion particle geometries by proposing a shape factor. On the basis of this model, when the larger thermal conductivity of the nanoparticles is 100 times higher than the thermal conductivity of the base fluids, the thermal conductivity can be written as:

$$k_{nf} = k_f \left( \frac{k_S + (m+1)k_f - (m+1)(k_f - k_S)\phi}{k_S + (m+1)k_f + (k_f - k_S)\phi} \right), \qquad (16)$$

where subscripts $f$ and $S$ denotes the base fluid phase and nanoparticles phase, $k_f$ and $k_S$ are the thermal conductivities of the base fluid and nanoparticle, respectively. $m$ is the shape factor of the Hamilton-Crosser model [24] and the shape factor values of different shapes are noted in Table 1. The shape factor should be noted $m = \frac{3}{\lambda}$, here $\lambda$ is the sphericity (the ratio between the surface areas of sphere and real particles with equal volumes). The sphericity of the needle and lamina are 0.62 and 0.185, respectively. The shape factor of the particle is 3 $(m = 3)$, In this case, Hamilton-Crosser model becomes a Maxwell- Garnett model. Thermophysical properties of Copper (II) oxide are mentioned in Table 2.

The wave and laboratory frames are introduced through

$$\hat{x} = \hat{X} - s\hat{t}, \qquad \hat{u} = \hat{U} - s, \qquad \hat{y} = \hat{Y}, \qquad \hat{v} = \hat{V}. \qquad (17)$$

The corresponding dimensional boundary conditions are

$$\left.\begin{array}{l} \hat{\psi} = \dfrac{q}{2}, \hat{u} = \dfrac{\partial \hat{\psi}}{\partial \hat{Y}} = -s - \hat{\beta}_1 \left(1 + \dfrac{1}{\beta}\right) \dfrac{\partial^2 \hat{\psi}}{\partial \hat{Y}^2}, \hat{T} + \hat{\beta}_2 \dfrac{\partial \hat{T}}{\partial \hat{Y}} = \hat{T}_0, \hat{C} + \hat{\beta}_3 \dfrac{\partial \hat{C}}{\partial \hat{Y}} = \hat{C}_0 \text{ at } \hat{Y} = \hat{h}_1, \\[2mm] \hat{\psi} = -\dfrac{q}{2}, \hat{u} = \dfrac{\partial \hat{\psi}}{\partial \hat{Y}} = -s + \hat{\beta}_1 \left(1 + \dfrac{1}{\beta}\right) \dfrac{\partial^2 \hat{\psi}}{\partial \hat{Y}^2}, \hat{T} - \hat{\beta}_2 \dfrac{\partial \hat{T}}{\partial \hat{Y}} = \hat{T}_1, \hat{C} - \hat{\beta}_3 \dfrac{\partial \hat{C}}{\partial \hat{Y}} = \hat{C}_1 \text{ at } \hat{Y} = \hat{h}_2, \end{array}\right\} \qquad (18)$$

The following non-dimensional parameters are

$$\psi = \frac{\hat{\psi}}{sd_1},\ X = \frac{\hat{X}}{\lambda},\ Y = \frac{\hat{Y}}{d_1},\ t = \frac{s\hat{t}}{\lambda},\ x = \frac{\hat{x}}{\lambda},\ \delta = \frac{d_1}{\lambda},\ y = \frac{\hat{y}}{d_1},\ u = \frac{\hat{u}}{s},\ \theta = \frac{\hat{T}-\hat{T}_0}{\hat{T}_1-\hat{T}_0},$$

$$\gamma = \frac{\hat{C}-\hat{C}_0}{\hat{C}_1-\hat{C}_0},\ h_1 = \frac{\hat{h}_1}{d_1},\ h_2 = \frac{\hat{h}_2}{d_1},\ d = \frac{d_2}{d_1},\ a = \frac{a_1}{d_1},\ b = \frac{b_1}{d_1},\ v = -\delta\frac{\partial\psi}{\partial x},\ u = \frac{\partial\psi}{\partial y},$$

$$F = \frac{q}{sd_1},\ p = \frac{\hat{p}d_1^2}{s\lambda\mu_f},\ \beta_1 = \frac{\hat{\beta}_1}{d_1},\ \beta_2 = \frac{\hat{\beta}_2}{d_1},\ \beta_3 = \frac{\hat{\beta}_3}{d_1},\ Gr = \frac{g(\rho\beta_T)_{nf} d_1^2(\hat{T}_1-\hat{T}_0)}{s\mu_f},$$

$$Br = \frac{g(\rho\beta_C)_{nf} d_1^2(\hat{C}_1-\hat{C}_0)}{s\mu_f},\ v = \frac{\hat{v}}{s},\ Sr = \frac{D_{nf}k_T(\hat{C}_1-\hat{C}_0)}{k_f(\hat{T}_1-\hat{T}_0)},\ Nb = \frac{(\rho c_P)_{nf} D_{Cf}(\hat{C}_1-\hat{C}_0)}{(\rho c_f)_{nf}\mu_f},$$

$$Nt = \frac{(\rho c_P)_{nf} D_{Tf}(\hat{T}_1-\hat{T}_0)}{(\rho c_f)_{nf}\mu_f T_m},\ \mathrm{Re} = \frac{\rho_f sd_1}{\mu_f},\ Q_1 = \frac{Q_0 d_1^2}{(\hat{T}_1-\hat{T}_0)k_f}. \qquad (19)$$

By using the non-dimensional terms then the equations (8)-(11) can be written as

$$\frac{\partial p}{\partial x} = \frac{1}{(1-\phi)^{2.5}}\left(1+\frac{1}{\beta}\right)\frac{\partial^2 u}{\partial y^2} + Gr\theta + Br\gamma, \qquad (20)$$

$$\frac{\partial p}{\partial y} = 0, \qquad (21)$$

$$\frac{1}{(1-\phi)^{2.5}}\left(1+\frac{1}{\beta}\right)\frac{\partial^4\psi}{\partial y^4} + Gr\frac{\partial\theta}{\partial y} + Br\frac{\partial\gamma}{\partial y} = 0, \qquad (22)$$

$$A^*\frac{\partial^2\theta}{\partial y^2} + Sr\frac{\partial^2\gamma}{\partial y^2} + Q_1 = 0, \qquad (23)$$

$$\frac{\partial^2\gamma}{\partial y^2} + \frac{Nt}{Nb}\frac{\partial^2\theta}{\partial y^2} = 0, \qquad (24)$$

where $A^* = \left[\dfrac{K_s + (m+1)K_f - (m+1)(K_f - K_s)\phi}{K_s + (m+1)K_f + (K_f - K_s)\phi}\right].$

The corresponding dimensionless boundary conditions are

$$\left.\begin{array}{l}\psi = \dfrac{F}{2},\ u = \dfrac{\partial\psi}{\partial y} = -1-\beta_1\left(1+\dfrac{1}{\beta}\right)\dfrac{\partial^2\psi}{\partial y^2},\ \theta+\beta_2\dfrac{\partial\theta}{\partial y} = 0,\ \gamma+\beta_3\dfrac{\partial\gamma}{\partial y} = 0\ \text{at}\ y = h_1 = a\cos[2\pi(x-t)]+1,\\[6pt] \psi = -\dfrac{F}{2},\ u = \dfrac{\partial\psi}{\partial y} = -1+\beta_1\left(1+\dfrac{1}{\beta}\right)\dfrac{\partial^2\psi}{\partial y^2},\ \theta-\beta_2\dfrac{\partial\theta}{\partial y} = 1,\ \gamma-\beta_3\dfrac{\partial\gamma}{\partial y} = 1\ \text{at}\ y = h_2 = -b\cos[2\pi(x-t)+\omega]-d.\end{array}\right\} \qquad (25)$$

Here we have considering $F$ is the mean flow over a period is

$$\Theta = F + 1, \qquad F = \int_{h_1}^{h_2} \frac{\partial \psi}{\partial y} dy, \qquad (26)$$

where $\Theta = \frac{Q}{sd_1}$ and $F = \frac{q}{sd_1}$.

**Table 1: Shape of the nanoparticles with their shape factors.**

| Nanoparticle type | Shape | Shape factor | Sphericity |
|---|---|---|---|
| Needle | 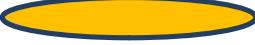 | 4.9 | 0.62 |
| Lamina | 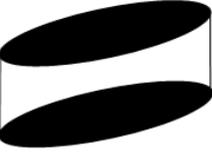 | 16.2 | 0.185 |

**Table 2: Thermophysical properties of the nanofluid (CuO-water)**

| Properties | CuO | Base fluid (water) |
|---|---|---|
| Appearance | Black to Brown powder | White crystalline solid |
| Density $(\rho)$ | 6.31 g/cm$^3$ | 997.1 |
| Heat Capacity $(\rho C)_P$ | 531.80 | 4179.0 |
| Thermal Conductivity $k$ | 76.50 | 0.6130 |
| Prandtl Number $(\Pr)$ | - | 6.20 |
| Melting Point | 1,201° C (2,194° F) | 0.00°C (32.00° F, 273.15K) |
| Boiling Point | 2,000° C (3,632° F) | 99.98° C (211.96° F, 373.13K) |

## 3. Method of Solution

Dimensionless equations (22)-(24) are solved by using Homotopy Analysis Method (HAM), then the initial approximations and equivalent auxiliary linear operator are obtained as

$$\psi_0(y) = \beta_1\left(1+\frac{1}{\beta}\right)\left[\begin{array}{l}\dfrac{F}{2}\dfrac{y^3}{6}-\left(\dfrac{h_1+h_2}{2}\right)\dfrac{y^2}{2}+\left(\left(\dfrac{h_1+h_2}{4}\right)-\left(\dfrac{h_1^2+h_2^2}{4}\right)\right)y \\ +\left(\dfrac{h_1^3+h_2^3}{12}\right)+\left(\dfrac{(h_1+h_2)(h_1^2+h_2^2)}{8}\right)- \\ \left(\dfrac{h_1+h_2}{8}\right)\left((h_1+h_2)-(h_1^2+h_2^2)\right)\end{array}\right], \quad (27)$$

$$\theta_0(y) = \frac{2}{\beta_2}\left[\frac{h_1+h_2}{2}-y\right], \quad (28)$$

$$\gamma_0(y) = \frac{2}{\beta_3}\left[\frac{h_1+h_2}{2}-y\right]. \quad (29)$$

$$L_\psi = \frac{\partial^4}{\partial y^4}, \qquad L_\theta = \frac{\partial^2}{\partial y^2}, \qquad L_\gamma = \frac{\partial^2}{\partial y^2}. \quad (30)$$

**Zeroth-order deformation equations are**

$$(1-q)L_\psi\left[\psi(y,q)-\psi_0(y,q)\right] = qH_\psi h_\psi N_\psi\left[\psi(y,q),\theta(y,q),\gamma(y,q)\right], \quad (31)$$

$$(1-q)L_\theta\left[\theta(y,q)-\theta_0(y,q)\right] = qH_\theta h_\theta N_\theta\left[\theta(y,q),\gamma(y,q)\right], \quad (32)$$

$$(1-q)L_\gamma\left[\gamma(y,q)-\gamma_0(y,q)\right] = qH_\gamma h_\gamma N_\gamma\left[\gamma(y,q),\theta(y,q)\right], \quad (33)$$

in the above equations, $h_\psi$, $h_\theta$ and $h_\gamma$ are the auxiliary parameters, the embedded parameter is $q\in[0,1]$, auxiliary linear operator $L$, $H_\psi$, $H_\theta$ and $H_\gamma$ are the auxiliary functions. Nonlinear operators $N_\psi, N_\theta$ and $N_\gamma$ can be written as

$$N_\psi\left[\psi(y,q),\theta(y,q),\gamma(y,q)\right] = \frac{1}{(1-\phi)^{2.5}}\left(1+\frac{1}{\beta}\right)\frac{\partial^4\psi(y,q)}{\partial y^4}+Gr\frac{\partial\theta(y,q)}{\partial y}+Br\frac{\partial\gamma(y,q)}{\partial y}, \quad (34)$$

$$N_\theta\left[\theta(y,q),\gamma(y,q)\right] = A*\frac{\partial^2\theta(y,q)}{\partial y^2}+Sr\frac{\partial^2\gamma(y,q)}{\partial y^2}+\Pr Q_1, \quad (35)$$

$$N_\gamma\left[\gamma(y,q),\theta(y,q)\right] = \frac{\partial^2\gamma(y,q)}{\partial y^2}+\frac{Nt}{Nb}\frac{\partial^2\theta(y,q)}{\partial y^2}. \quad (36)$$

The initial approximations $\psi_0(y,q)$, $\theta_0(y,q)$ and $\gamma_0(y,q)$ approaches to $\psi(y,q)$, $\theta(y,q)$ and $\gamma(y,q)$, respectively. The $q$ values taken from 0 to 1. Mathematically,

$$\begin{array}{l}\psi(y,0)=\psi_0(y), \ \theta(y,0)=\theta_0(y), \ \gamma(y,0)=\gamma_0(y) \\ \psi(y,1)=\psi(y), \quad \theta(y,1)=\theta(y), \quad \gamma(y,1)=\gamma(y)\end{array} \quad (37)$$

**n-order deformation equations of the given problem are**

$$L_\psi\left[\psi_n(y)-\xi_n\psi_{n-1}(y)\right] = h_\psi R_n^\psi(y), \quad (38)$$

$$L_\theta\left[\theta_n(y) - \xi_n \theta_{n-1}(y)\right] = h_\theta R_n^\theta(y), \tag{39}$$

$$L_\gamma\left[\gamma_n(y) - \xi_n \gamma_{n-1}(y)\right] = h_\gamma R_n^\gamma(y). \tag{40}$$

with

$$R_n^\psi(y) = \frac{1}{(n-1)!} \frac{\partial^{n-1} N_\psi\left[\psi(y,q), \theta(y,q), \gamma(y,q)\right]}{\partial q^{n-1}}\bigg|_{q=0}, \tag{41}$$

$$R_n^\theta(y) = \frac{1}{(n-1)!} \frac{\partial^{n-1} N_\theta\left[\theta(y,q), \gamma(y,q)\right]}{\partial q^{n-1}}\bigg|_{q=0}, \tag{42}$$

$$R_n^\gamma(y) = \frac{1}{(n-1)!} \frac{\partial^{n-1} N_\gamma\left[\gamma(y,q), \theta(y,q)\right]}{\partial q^{n-1}}\bigg|_{q=0}, \tag{43}$$

where

$$\xi_n = \begin{cases} 0, & n \leq 1 \\ 1, & n \succ 1. \end{cases} \tag{44}$$

## Convergence Analysis

In this section the convergence of Homotopy Analysis Method (HAM) are stated. $\psi(h_1)$, $\psi(h_2)$, $\theta(h_1)$, $\gamma(h_1)$ and $\gamma(h_2)$ are contains the auxiliary parameters $h_\psi, h_\theta$ and $h_\gamma$. Frame the $\hbar$- curves at 35$^{th}$ order approximation to find the proper values of $h_\psi, h_\theta$ and $h_\gamma$ (see in figure 1(a & b)). The auxiliary linear parameters are adjusting and controlling the homotopic solutions. In this problem, we have chosen the convergence solution as $h_\psi = h_\theta = h_\gamma = -0.7$.

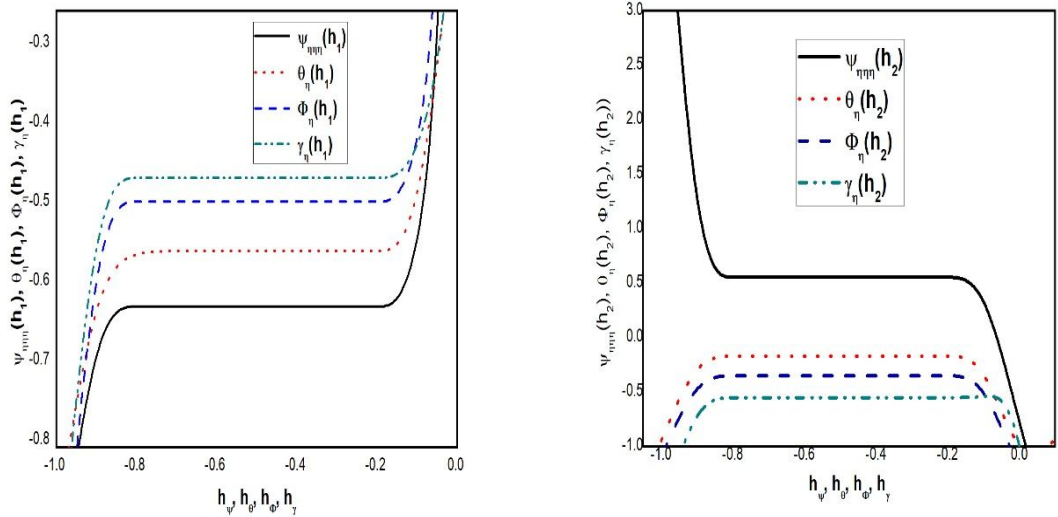

**Figures 1(a) and 1(b):** $\hbar$ curves for the function of $\psi(h_1)$, $\psi(h_2)$, $\theta(h_1)$, $\theta(h_2)$, $\gamma(h_1)$ and $\gamma(h_2)$ at $35^{th}$ order approximations when $b = 0.5$, $t = 0.2$, $a = 0.5$, $x = 0.2$, $\beta_1 = 1.2$, $d = 1.5$, $\phi = 0.7$, $Br = 0.8$, $Gr = 0.8$, $Nt = 0.6$, $N_{TC} = 0.5$, $Q_1 = 1.5$, $\beta = 0.8$, , $\beta_2 = 1.2$, $\beta_3 = 1.2$, $Nb = 0.6$.

## 4. Results and Discussion

This part of the research paper aims to discuss the velocity, thermal energy and concentration of the nanoparticles via graphs. The thermophysical properties of copper (II) oxide and water are mention in Table 2 and the different nanoparticle shapes and shape factor are noted in Table 1.

The impacts of physical parameters likeNanoparticle Grashof number $Br$, joule heating effect,velocity slip effect $\beta_1$, Thermal Grashof number $Gr$, thermophoresis diffusion parameter $Nt$, Casson fluid parameter $\beta$, Soret number $Sr$, Brownian motion parameter $Nb$, thermal slip parameter $\beta_2$, nanoparticle slip parameter $\beta_3$ on velocity, temperatureand nanoparticle concentration of cupric oxide nanoparticles containing different shapes of nanoparticles, i.e., needleand lamina are analyzed. The influence of physical parameters is represented by the graphs in various colors. Further, the increase or decrease in physical quantities for various fluid flow parameters is shown by the direction of arrows.

### 4.1. Velocity Distribution

The influence of several flow parameters on velocity field are show in figs. 2 (a)-(d). These figures observed that velocity field traces a parabolic trajectory with the middle of channel is maximum. Figure 2(a) exhibits the velocity profile for various values of Casson fluid $\beta$. The Casson fluid parameter $\beta$ increases with an enhancing the velocity profile. Physically, rising values of $\beta$ develop the viscous forces. These viscous forces have propensity to reduce the thermal boundary layer. In the figure 2(b) shows that the velocity slip parameter $\beta_1$ on velocity. It is found the velocityslip parameter $\beta_1$ increase with an increasing the velocity field. Figure 2(c) depicts the impact of thermal Grashof number $Gr$ on the velocity field. Form the figure 2(c), we have observed that the velocity enhancedwith higher values of $Gr$. Physically, the buoyancy force of the fluid flow can be increases to the higher values thermal Grashof number because free convection influences. Figure 2(d) examines the influence of nanoparticle Grashof number $Br$ on velocity. We have observed increasing the nanoparticle Grashof number then the velocity is enhanced.

### 4.2. Temperature profile

Figures 3(a)-(e) show variation of temperature profile as a function of y. Here variations of joule heating effect, thermophoresis parameter $Nt$, thermal slip parameter $\beta_2$, soret number and Brownian motion paramter $Nb$ are seen.

Figure 3(a) shows the impact of thermal slip parameter $\beta_2$ on the thermal energy. The thermal energy enhanced with an increasing thermal slip parameter $\beta_2$. Figure 3(b) depicts the effect of soret number on thermal energy, here higher values of soret number with an increases the thermal energy. Figure 3(c) and figure 3(d) examines the temperature profile with various values of thermophoresis and Brownian motion paramters. These figures examines the therma enenrgy increases with an increasing Brownian motion parameter but opposite behaviour of the thermohoresis parameter and the physical reson is that the Brownian motion gives rise to the nanoparticles to rearrange, forming a blend, which enhancing the thermal conductivity. In the figure 3(e) examine the joule heating effect on temperature profile. It is found the joule heating effect increases with an increasing the temperature profile.

### 4.3. Nanoparticle concentration profile

Figures 4 (a)-(d) show the behavior of nanoparticle concentration for the variation of joule heating effect, thermal slip parameter $\beta_2$, soret number, thermophoresis parameter and Brownian motion paramter.

Figure 4 (a) analyze the influence of concentration slip parameter $\beta_2$ on nanoparticle concentration. The results show that decreasing in temperature profile presences in a specific domain with increasing the Slip parameter $\beta_2$. After that domain, the relationship is reflected with the Slip parameter $\beta_2$. The figures 4 (b) and 4 (c) are examines the influence of the $Nb$ and $Nt$ on the nanoparticle concentration. It is examining that by increasing $Nb$, the concentration of nanoparticle decreases. However, an opposite trend has been observed as $Nt$. This is because, the random motion of nanoparticles getting increased with an increasing $Nb$, which in turn an enhancement of fluid temperature and reduction of the nanoparticle diffusion.

It is noticed from figure 4 (d) examine joule heating on nanoparticle concentration profile, here higher values of joule heating effect with an decreases the nanoparticle concentration.

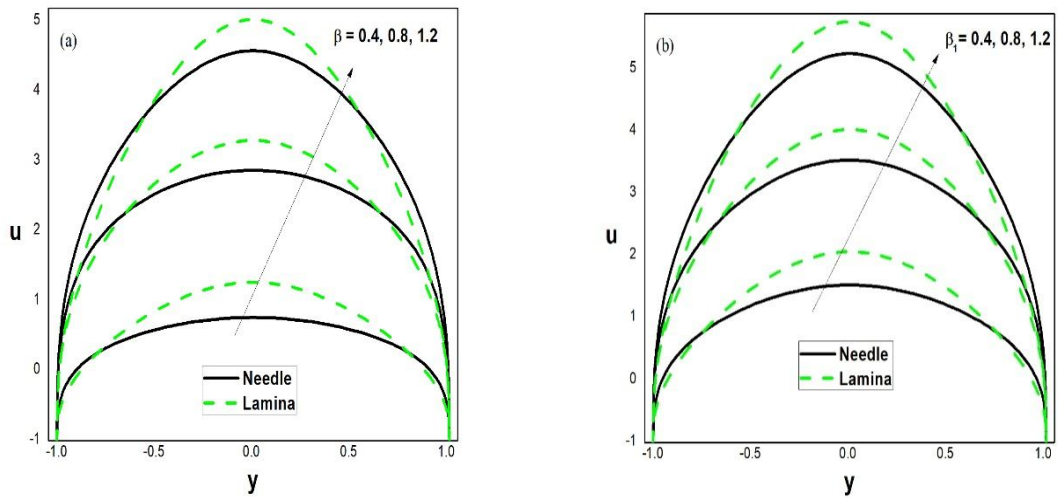

**Figures2(a)-(b):** variation of velocity profile versus y. when $b = 0.5$, $t = 0.2$, $x = 0.2$, $a = 0.5$, $d = 1.5$, $Sr = 0.8$, $Nt = 0.6$, $Q_1 = 1.5$, $Nb = 0.6$, $\beta_2 = 1.2$, $\beta_3 = 1.2$ and $\omega = 0.7$. (a) $Br = 0.8$, $Gr = 0.8$ and $\beta_1 = 1.2$. (b) $Br = 0.8$, $Gr = 0.8$ and $\beta = 0.7$.

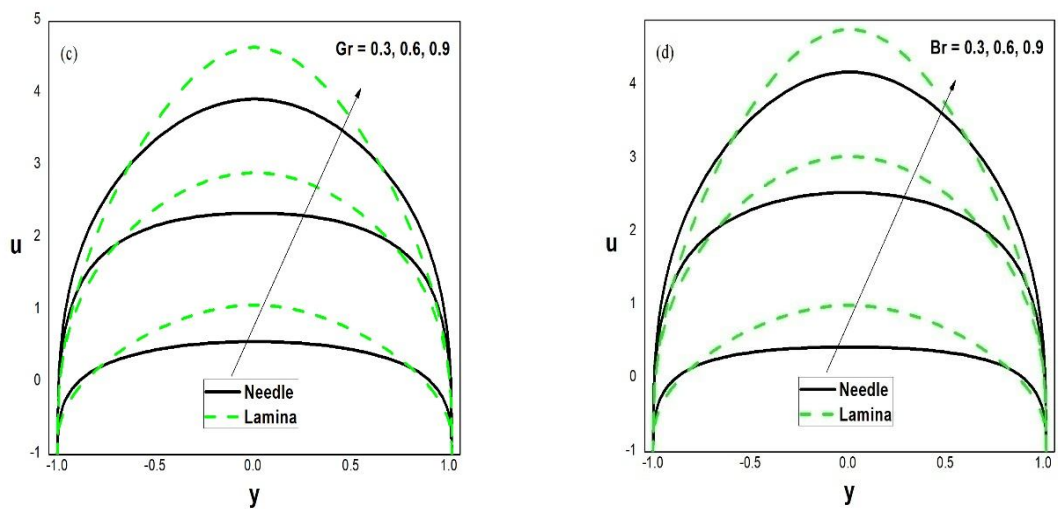

**Figures2(c)-(d):** variation of velocity profile versus y. when $b = 0.5$, $t = 0.2$, $x = 0.2$, $a = 0.5$, $d = 1.5$, $Sr = 0.8$, $Nt = 0.6$, $Q_1 = 1.5$, $Nb = 0.6$, $\beta_2 = 1.2$, $\beta_3 = 1.2$ and $\omega = 0.7$. (c) $Br = 0.8$, $\beta = 0.7$ and $\beta_1 = 1.2$. (d) $Gr = 0.8$, $\beta = 0.7$, $\beta_1 = 1.2$.

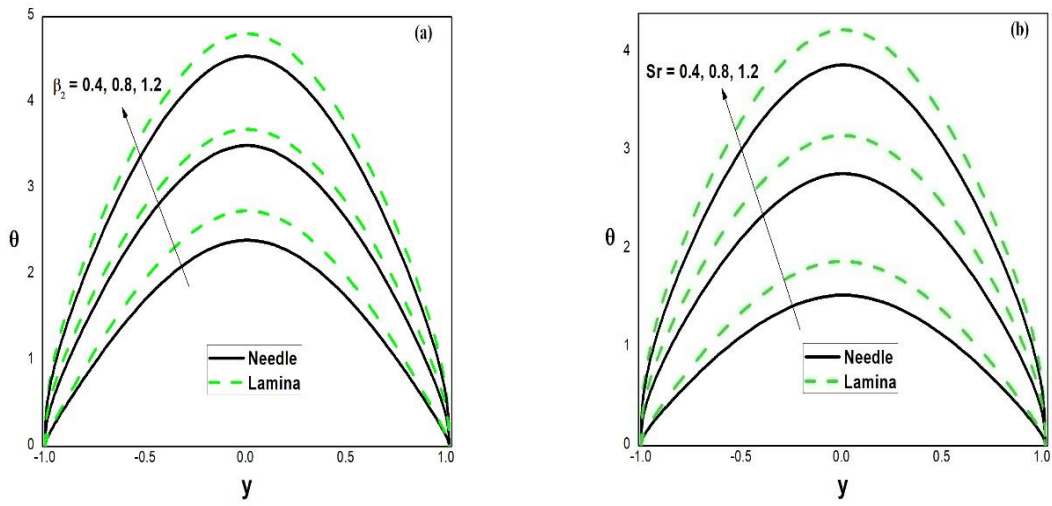

**Figures 3(a)-(b):** variation of temperature profile versus y. when $b = 0.5$, $t = 0.2$, $x = 0.2$, $a = 0.5$, $\beta_3 = 1.2$, $d = 1.5$ and $\omega = 0.7$. (a) $Nt = 0.6$, $Q_1 = 1.5$, $Nb = 0.6$ and $Sr = 0.8$. (b) $Nt = 0.6$, $Q_1 = 1.5$, $Nb = 0.6$ and $\beta_2 = 1.2$.

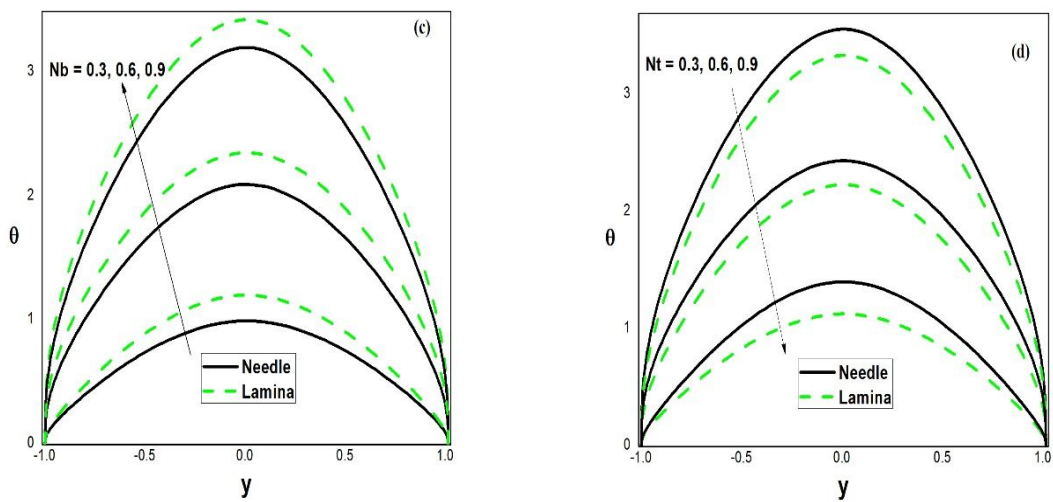

**Figures 3(c)-(d):** variation of temperature profile versus y. when $b = 0.5$, $t = 0.2$, $x = 0.2$, $a = 0.5$, $\beta_3 = 1.2$, $d = 1.5$ and $\omega = 0.7$. (c) $Nt = 0.6$, $Q_1 = 1.5$, $\beta_2 = 1.2$ and $Sr = 0.8$. (d) $\beta_2 = 1.2$, $Q_1 = 1.5$, $Nb = 0.6$ and $Sr = 0.8$.

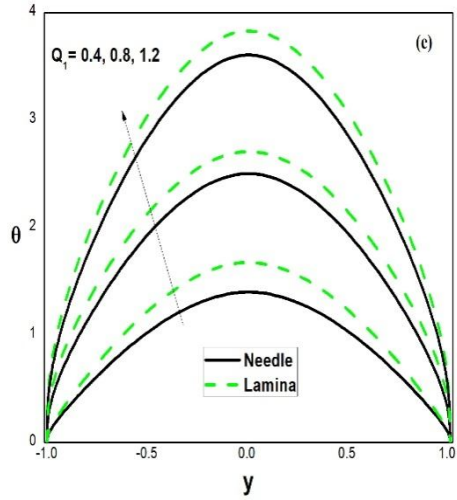

**Figure 3(e):** variation of temperature profile versus y. when $b = 0.5$, $t = 0.2$, $x = 0.2$, $a = 0.5$, $\beta_3 = 1.2$, $d = 1.5$ and $\omega = 0.7$. (e) $Nt = 0.6$, $\beta_2 = 1.2$, $Nb = 0.6$ and $Sr = 0.8$.

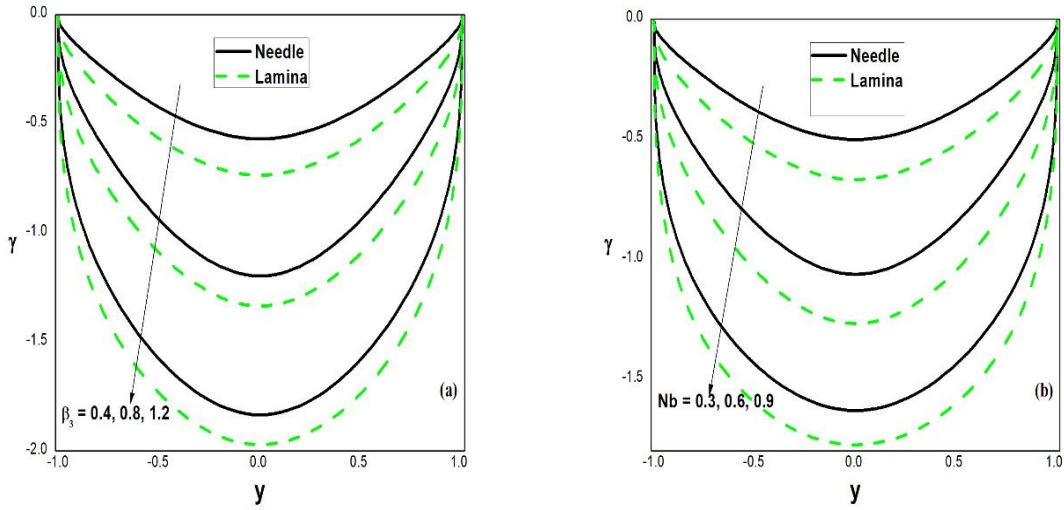

**Figures 4(a)-(b)** variation of nanoparticle concentration versus y. when $b = 0.5$, $t = 0.2$, $x = 0.2$, $Sr = 0.8$, $a = 0.5$, $\beta_2 = 1.2$, $d = 1.5$ and $\omega = 0.7$. (a) $Nt = 0.6$, $Nb = 0.6$ and $Q_1 = 1.5$. (b) $\beta_3 = 1.2$ $Nt = 0.6$ and $Q_1 = 1.5$.

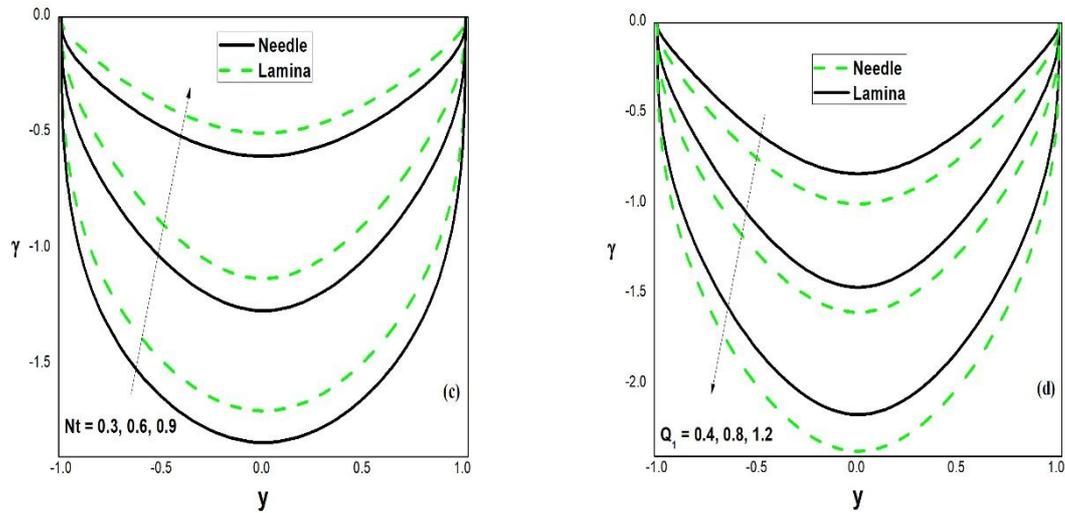

**Figures4(c)-(d):** variation of nanoparticle concentration versus y. when $b = 0.5$, $t = 0.2$, $x = 0.2$, $Sr = 0.8$, $a = 0.5$, $\beta_2 = 1.2$, $d = 1.5$ and $\omega = 0.7$. (c) $\beta_3 = 1.2$, $Nb = 0.6$ and $Q_1 = 1.5$. (d) $Nt = 0.6$, $Nb = 0.6$ and $\beta_3 = 1.2$.

## Conclusion

In the present research paper, joule heating effect on peristaltic transport of a slip condition with two different shapes of nanoparticles such as needle and lamina have been analyzed. The following important points observations were established:

- ➢ Increases in joule heating parameter enhances the fluid temperature and nanoparticle concentration decreases with an increasing joule heating effect.

- ➢ Increasing the velocity slip parameter generally accelerates the velocity of the fluid flow enhanced in different shapes at the middle of the channel and the ends of the channel walls is reduced and lamina shape of the Copper (II) oxide water is greater than needle shape. In the presence of fluid with nanoparticles, therefore assist the fluid flow.

- ➢ Both thermal and nanoparticle Grashof numbers enhance the velocity profile.

- ➢ The small change for the nanofluids containing nanoparticles shape needle is less observed as compared to containing nanoparticle shapelamina.The thermal conductivity of the nanoparticles is completely accordance with expectation, the nanoparticles shape needle has very low thermal conductivity as compared to nanoparticleshape lamina.